\documentstyle[amssymb,preprint,aps]{revtex}
\begin{document}
\title{Finite-wavevector phonon coupling to degenerate electronic states in La$%
_{2-x}$Sr$_{x}$CuO$_{4}.$}
\author{V.V.Kabanov and D.Mihailovic}
\address{Jozef Stefan Institute, Jamova 39, 1000 Ljubljana, Slovenia}
\maketitle

\begin{abstract}
We consider finite-$k$ electron-phonon coupling appropriate for a
superconductor with a very short coherence length and local intersite pairs.
From group theoretical analyis we find that $k\neq 0\,\ $phonons can couple
to degenerate electronic states in La$_{2-x}$Sr$_{x}$CuO$_{4}$ and a
Jahn-Teller-like distortion is found to be operative for $E_{u}$ and $E_{g}$%
-symmetry O states. Experimentally observed anomalies in inelastic neutron
scattering, ESR\ and EXAFS are found to be consistent with such an
interaction of the $\tau _{1}$ symmetry $\Sigma -\,$point mode with the
planar O p$_{x}$ and p$_{y}$ states. The proposed interaction is found to
naturally facilitate symmetry breaking associated with pairing and/or stripe
formation.
\end{abstract}

\date{}

\newpage

\section{Introduction}

The existence of local intersite pairs in cuprate superconductors is
inferred from the very short coherence lengths in these materials, which is
in turn determined from measurements of $H_{c2}$. Given that the pair
dimensions $d$ cannot exceed the coherence length, i.e. $d\lesssim \xi $, we
may infer that any possible lattice distortions associated with pairing have
a finite range, defined by the extent of the pair $\sim d$. However, the
particular form of the electron-phonon interaction leading to these
distortions and particular type of phonon mode responsible for pairing
should be determined from experiments. According to inelastic neutron
scattering data a large (30\%) anomaly in $\tau _{1}$ phonon mode
corresponding to the ($\zeta ,0,0$) direction has been observed in a number
of doped cuprates (but not in the parent compound)\cite{Egami}.

The observed anomaly takes place at an incomensurate wavevector and is
suggested to be associated with pairing and/or stripe formation. The
electron phonon interaction can then be written in the form: 
\begin{equation}
g(k_{0},{\bf k})=g_{0}/((k-k_{0})^{2}+\gamma ^{2})
\end{equation}
where $g_{0}$ is a constant describing the strength of coupling, $k_{0}\,$\
defines the wavevector asociated with the interaction and its range in $k$%
-space, while $\gamma $ defines its width. In this paper we analyse the $e-p$
coupling of phonon modes with $k\neq 0\,\ $using group theory applicable to
La$_{2-x}$Sr$_{x}$CuO$_{4}$ and discuss the relevant phonon modes by
examining the available experimental data from neutron diffraction, ESR and
EXAFS. We find that the experimentally observed features can be described by
the coupling of a $\tau _{1}$ - symmetry in-plane O-mode along the $\Sigma $
direction to holes occupying $E_{g}$ or $E_{u}$ - symmetry planar O states.

\section{E-p coupling}

The Brillouin zone (BZ) corresponding to the tetragonal point group $D_{4h}$
applicable for La$_{2-x}$Sr$_{x}$CuO$_{4}$ is shown in Figure 1. To consider
local pairs and/or stripes forming along the Cu-O bond direction or along 45$%
^{\circ }$ to it, we need to consider the $\Sigma $ and the $\Delta $
points, corresponding to the ($\zeta ,0,0)$ and ($\zeta ,\zeta ,0)$
directions respectively. The special symmetry points ($\Gamma ,X$ and $M$
etc.) give rise to commensurate stripes\cite{1/8}.

Before proceeding with analysis of the e-p coupling for the case of general $%
k$, it should be pointed out since the symmetrised cross product of the
representations at the $\Gamma $ point, 
\begin{equation}
\lbrack E_{u}\times E_{u}]=[E_{g}\times E_{g}]=A_{1g}+B_{1g}+B_{2g}
\end{equation}
the electrons can couple only with $A_{1g}$ modes (there are no $B_{1g}$ and 
$B_{2g}$ at the $\Gamma $ point of La$_{2-x}$Sr$_{x}$CuO$_{4}$). (There are
two such modes in $D_{4h}$ and are associated either with apex oxygen O(2)
ions or La ions.) These do not lead to any symmetry breaking. It follows
that stripe formation or further symmetry breaking cannot be associated with
the $\Gamma $ point ($k=0)$, but can only be {\em associated with finite }$%
{\bf k}$. For the case of tetragonal symmetry, this corresponds to the $%
\Sigma $ and the $\Delta $ points in the BZ.

\subsection{Coupling to non-degenerate states}

Since the $\Sigma $ and $\Delta $ points have a four pronged star, we
propose the following form of coupling to electrons in single non-degenerate
electronic states: 
\begin{equation}
H_{int}=\sum_{{\bf l},s}n_{{\bf l},s}\sum_{k_{0}=1}^{4}\sum_{{\bf k}}g(k_{0},%
{\bf k})\exp {(i{\bf kl})}(b_{-{\bf k}}^{\dagger }+b_{{\bf k}})
\end{equation}
where ${\bf l}$ is the site label and 
\begin{equation}
g(k_{0},{\bf k})=g/((k-k_{0})^{2}+\gamma ^{2})
\end{equation}
where $k_{0}$ are the 4 wavevectors corresponding to the prongs of the star
associated with stripe (pair) formation. We should stress that in this case
the nondegenerate electronic states are associated with $p_{z}$-orbitals of
planar oxygens which transform as $A_{2u}$ or $B_{2u}$. The Hamiltonian
above on its own does not lead to symmetry breaking, but phase separation
can in principle still take place as a result of competition of short-range
attraction and long-range Coulomb repulsion forces. However, the symmetry of
both phases would be the same.

\subsection{Coupling to degenerate states (Jahn-Teller-like coupling)}

A more interesting case arises when two-fold degenerate levels (for example
the two $E_{u}$ states corresponding to the planar O $p_{x}$ and $p_{y}$
orbitals or the $E_{u}$ and $E_{g}$ states of the apical O) interacts with $k%
\not=0$ phonons. Again, we are interested in the phonons which lead to
symmetry breaking and allow the formation of stripes. The invariant form of
Hamiltonian has a simple form for phonons which transforms as the $\tau _{1}$
representations of the group of wave-vector $G_{k}$. Taking into account
that $E_{g}$ and $E_{u}$ representations are real and Pauli matrices $\sigma
_{i}$ corresponding to the doublet of $E_{g}$ or $E_{u}$ transform as $A_{1g}
$ ($k_{x}^{2}+k_{y}^{2}$) for $\sigma _{0}=\left( 
\begin{array}{cc}
1 & 0 \\ 
0 & 1
\end{array}
\right) $, $B_{1g}$ ($k_{x}^{2}-k_{y}^{2}$) for $\sigma _{3}=\left( 
\begin{array}{cc}
1 & 0 \\ 
0 & -1
\end{array}
\right) $, $B_{2g}$ ($k_{x}k_{y}$) for $\sigma _{1}=\left( 
\begin{array}{cc}
0 & 1 \\ 
1 & 0
\end{array}
\right) $, and $A_{2g}$ ($s_{z}$) \ for $\sigma _{2}=\left( 
\begin{array}{cc}
0 & i \\ 
-i & 0
\end{array}
\right) $representations respectively, we can construct an invariant form of
the Hamiltonian: 
\begin{eqnarray}
H_{int} &=&\sum_{{\bf l},s}\sigma _{0,{\bf l}}\sum_{k_{0}=1}^{4}\sum_{{\bf k}%
}g_{0}(k_{0},{\bf k})\exp {(i{\bf kl})}(b_{-{\bf k}}^{\dagger }+b_{{\bf k}})+
\nonumber \\
&&\sum_{{\bf l},s}\sigma _{3,{\bf l}}\sum_{k_{0}=1}^{4}\sum_{{\bf k}%
}g_{1}(k_{0},{\bf k})(k_{x}^{2}-k_{y}^{2})\exp {(i{\bf kl})}(b_{-{\bf k}%
}^{\dagger }+b_{{\bf k}})+ \\
&&\sum_{{\bf l},s}\sigma _{1,{\bf l}}\sum_{k_{0}=1}^{4}\sum_{{\bf k}%
}g_{2}(k_{0},{\bf k})k_{x}k_{y}\exp {(i{\bf kl})}(b_{-{\bf k}}^{\dagger }+b_{%
{\bf k}})  \nonumber \\
&&\sum_{{\bf l},s}\sigma _{2,{\bf l}}S_{z,{\bf l}}\sum_{k_{0}=1}^{4}\sum_{%
{\bf k}}g_{3}(k_{0},{\bf k})\exp {(i{\bf kl})}(b_{-{\bf k}}^{\dagger }+b_{%
{\bf k}})  \nonumber
\end{eqnarray}
where 
\begin{equation}
g_{i}(k_{0},{\bf k})=g_{i}/((k-k_{0})^{2}+\gamma ^{2})
\end{equation}

The first term in (5) describes the symmetric coupling and is identical to
the non-degenerate case. The second and third terms describe the e-p
interaction corresponding to the $\Sigma $ and $\Delta $ directions
respectively, while the last term describes the {\em coupling to spins}.

The proposed interaction (5) on its own results in splitting of the
degenerate states (a Jahn-Teller-like effect), breaking tetragonal symmetry
and resulting in a local orthorhombic distortion. It can therefore lead to
the formation of pairs and stripes with no further interaction. Of course
the stability and size of such a distortion will be determined by the
balance of short-range attraction and long-range Coulomb repulsion\cite
{Emery}.

\section{Discussion}

To determine the relevant wavevector $k_{0}$ applicable in (5) and (6) we
need to look at the experimental data. It is well known that for $k_{0}=0,$
there are no significant phonon anomalies (in Raman or infrared
spectroscopies for example) which can be associated either with $T_{c}$ or
the pseudogap $T_{p}$. However, local probe experiments like neutron
scattering\cite{Egami}, EXAFS\cite{Bianconi} and ESR\cite{Kochelaev} give a
different picture, and point to the existence of sizeable lattice anomalies
for $k\neq 0$. The lattice distortions associated with these experimental
observations are shown in Figure 2. Particularly the inelastic neutron data
(Fig. 2a)\ should be singled out here because of the unambiguity of the raw
data. In these experiments a large renormalisation of phonon frequency is
ubiquitously observed along the $\Sigma $ direction in many cuprates,
including 214 and 123 structure superconductors, but {\em not }in their
parent insulators. Moreover, its temperature dependence clearly suggests a
connection with pair formation\cite{Egami}.

The phonon $\tau _{1}$ modes corresponding to the $\Sigma $ direction which
can couple to the electrons are shown in Figure 3. One of these, O(1)1 can
be clearly identified as the mode observed in neutron scattering experiments 
\cite{Egami} and may thus be indentified as the mode responsible for pairing
or stripe formation. We also note that from the experiments, the width in $k$%
-space of the anomaly is of the same order as the magnitude$.$

The interaction given in Eqs.(5) and (6) leads directly to phase segregation
of distorted and undistorted domains, where the size of domains is
determined by the interplay of short-range attraction and screened Coulomb
repulsion as already mentioned\cite{Emery}. The incommensurability of the
interaction leads to modulation of charge density within the stripe.
Importantly since the number of charged carriers in the stripe is less than
the number of sites occupied by stripe, we may expect to observe itinerant
behaviour in such stripes, i.e. the charge carriers are\ not localised
within the stripe itself. At low doping when screening is reduced, the
formation of the stripes of the length 2 (pairs) is most probable, while at
lower temperatures and higher doping clearly longer stripes are more
probable. These pairs, contrary to the case of the Holstein interaction is
more extended in space and its size is governed by $k_{0}$.

\section{Conclusion}

The central idea that the {\em pair} is the ground state entity which
defines the extent of the e-p interaction leads to an interaction
Hamiltonian of degenerate modes with $k\neq 0$ phonons which is suggested to
describe the pairing and stripe formation in the cuprates. The splitting of
the degeneracy of planar O orbitals gives rise to a deformation either along
the bond axes or along the diagonals, depending on which mode is chosen, but
inelastic neutron scattering experiments suggest that the important
interaction is along the Cu-O bonds, giving rise to an $s+d$ - like form of
the interaction as well as coupling to spins via an $A_{2g}$-symmetry term.
Remarkably, the interaction which splits the degeneracy of the O states via
a JT-like effect is in spirit, if not in detail similar as was originally
suggested by Bednorz and Muller in their original paper on La$_{2-x}$Ba$_{x}$%
CuO$_{4}$\cite{BM}.

\section{Figure captions}

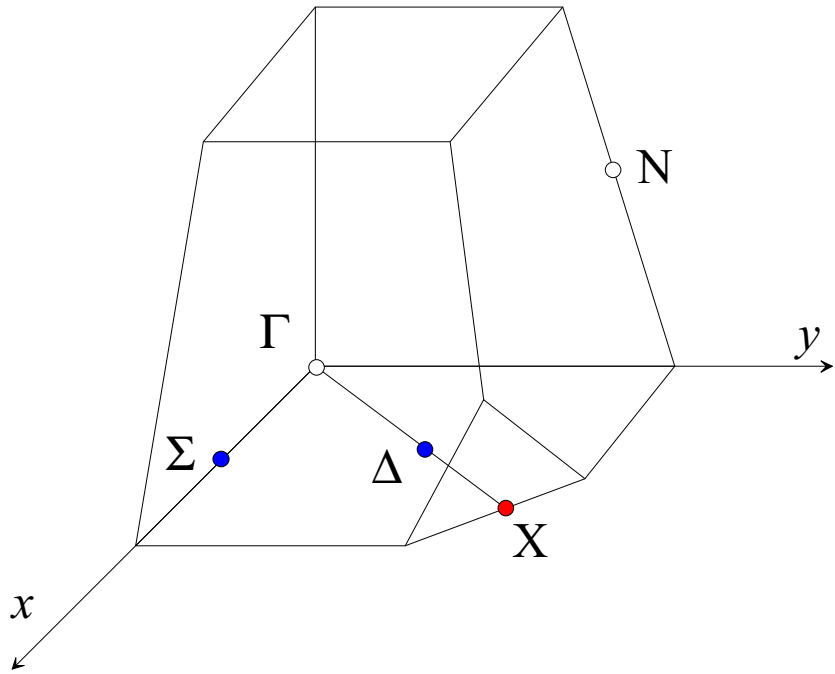
Figure 1. The Brillouin zone of the tetragonal structure corresponding to La$%
_{2-x}$Sr$_{x}$CuO$_{4}.$ According to experiment, the $\Sigma $ directions
is of relevance for pairing and stripe formation.

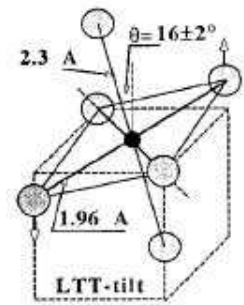
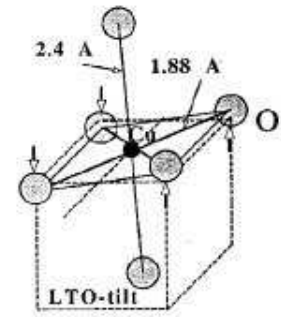
Figure 2. The vibrational modes causing the anomalies observed in a) neutron
scattering (ref. 1), b) ESR (ref. 4) and c)\ EXAFS (ref. 3)

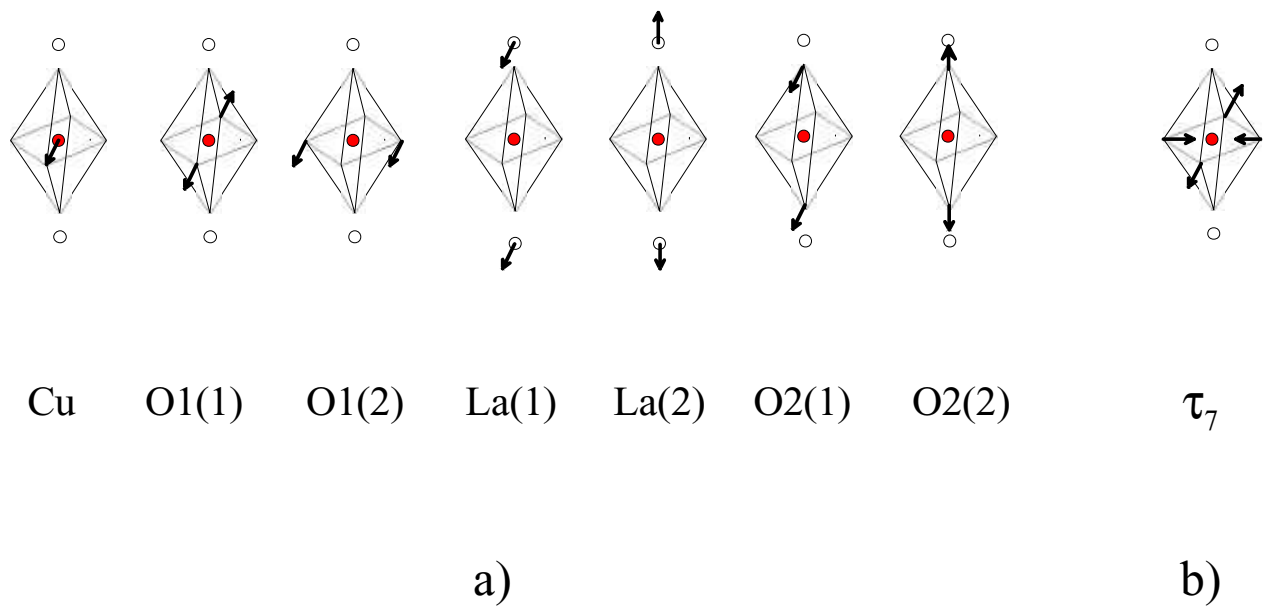
Figure 3. a) The wavevectors corresponding to the $\tau _{1}\;$%
representation at the $\Sigma -$point of the point group $D_{4h}$. Note that
the relative magnitude of the displacements shown depends on the value of $%
k_{0}$. b) The $\tau _{7}$ mode corresponding to the distortion suggested in
ref. 4 (Fig. 2 b) can also be viewed as a linear combination of two $\tau
_{1}$ modes at the $\Sigma -$point with different $k$.


\begin{references}
\bibitem{1/8}  The X and M points are relevant for the formation of
commensurate stripes near 1/8 doping.
\end{references}
\end{document}